\documentstyle[12pt,epsfig]{article} 
\begin{document}
\title{Generalized uncertainty principle and deformed dispersion relation induced by nonconformal metric flutuations}
\author{ A. Camacho
\thanks{email: acamacho@nuclear.inin.mx} \\
Department of Physics, \\
Instituto Nacional de Investigaciones Nucleares\\
Apartado Postal 18--1027, M\'exico, D. F., M\'exico.}
\date{}
\maketitle

\begin{abstract}
Considering the existence of nonconformal stochastic fluctuations in the metric tensor a generalized uncertainty principle and a deformed dispersion relation (associated to the propagation of photons) are deduced. Matching our model with the so called quantum $\kappa$--Poincar\'e group will allow us to deduce that the fluctuation--dissipation theorem could be fulfilled without needing a restoring mechanism associated with the intrinsic fluctuations of spacetime. In other words, the loss of quantum information is related to the fact that the spacetime symmetries are described by the quantum $\kappa$--Poincar\'e group, and not by the usual Poincar\'e symmetries. An upper bound for the free parameters of this model will also be obtained.
\end{abstract}
\bigskip
\bigskip


\section{Introduction}

Nowadays one of the most fascinating problems in modern physics is related to a, mathematically consistent, unified description of gravitation and quantum mechanics. Though, yet, there is no theory in this direction, it has been suggested that gravity should lead to a minimal observable distance [1, 2, 3]. This fact could have far--reaching consequences, for instance, it could imply ultraviolet regularization in field theory [4].

A second, and also very interesting prediction of these quantum gravity models comprises the so called deformed dispersion relation (which characterizes the propagation of massless particles) [5, 6, 7]. In connection with this last feature it has to be mentioned that if photons propagate with an energy dependent velocity, then this could imply the breakdown of Lorentz invariance [8].

In the present work it will be assumed that quantum gravity corrections appear as nonconformal stochastic fluctuations of the metric. It will be shown that these kind of fluctuations lead to the emergence of a deformed dispersion relation, for the propagation of photons, and that they imply the breakdown of Lorentz invariance. Additionally, it will be proved that they also render a modification of Heisenberg algebra. Confronting our model with the so called quantum $\kappa$--Poincar\'e group [3] will allow us to deduce that the fluctuation--dissipation theorem could be fulfilled without needing a restoring mechanism associated with the intrinsic fluctuations of spacetime [9]. We may rephrase this stating that the loss of quantum information is related to the fact that the spacetime symmetries are described by the quantum $\kappa$--Poincar\'e group, and not by the usual Poincar\'e symmetries. Finally, an upper bound for the free parameters of our model will be obtained.
\bigskip
\bigskip

\section{Nonconformal metric fluctuations and propagation of photons}
\bigskip

\subsection{Nonconformal fluctuations and white noise}
\bigskip 

Let us now suppose that the spacetime metric undergoes nonconformal stochastic fluctuations, and that these fluctuations represent white noise. This last condition seems to be a reasonable one [10], neverwithstanding, the introduction of color noise could be an interesting issue to analyze. In the context of decoherence effects, the possibilities that conformal flutuations offer have already been studied [10], and as will become clear below, no deformed dispersion relation, or modification to Heisenberg algebra, could emerge in this context. Hence, in order to have a richer spectrum of possibilities we introduce now nonconformal fluctuations of the background metric.

The nonconformal character implies that, from the outset, the maximal symmetry of the vacuum of the classical gravitational field will not be preserved. The loss of this characteristic allows us to introduce the so called quantum $\kappa$--Poincar\'e group [3] as part of our model, and therefore it will be possible to explain to absence of a restoring mechanism, associated with the fluctuations of spacetime, as a consequence of the loss of Lorentz invariance.

Hence, in the case where the average background metric is the Minkowkian one, we may write

{\setlength\arraycolsep{2pt}\begin{eqnarray}
ds^2 = e^{\psi(x)}\eta_{00}dt^2 + e^{\zeta(x)}\eta_{ij}dx^idx^j.
\end{eqnarray}}
\bigskip

Here we demand the following properties 

{\setlength\arraycolsep{2pt}\begin{eqnarray}
<e^{\psi(x)}\eta_{00}> = \eta_{00}, 
\end{eqnarray}}
 
{\setlength\arraycolsep{2pt}\begin{eqnarray}
<e^{\zeta(x)}\eta_{ij}> = \eta_{ij}. 
\end{eqnarray}}

From these last two expressions we conclude (assuming $\vert\psi(x)\vert<<1$) 

{\setlength\arraycolsep{2pt}\begin{eqnarray}
<\psi(x)> = 0, 
\end{eqnarray}}

{\setlength\arraycolsep{2pt}\begin{eqnarray}
<\partial_{\mu}\psi(x)> = 0. 
\end{eqnarray}}

These last conditions are related to the fact that $\psi(x)$ is white noise [10]. Of course, $\zeta(x)$ fulfills the same conditions. From (5) we find that $<\psi(x)^2> = cte.$, and if these fluctuations have a gaussian behavior, then

{\setlength\arraycolsep{2pt}\begin{eqnarray}
<\psi^2(x)> = \sigma_1^2, 
\end{eqnarray}}

{\setlength\arraycolsep{2pt}\begin{eqnarray}
<\zeta^2(x)> = \sigma_2^2, 
\end{eqnarray}}

\noindent $\sigma_1^2$ and $\sigma_2^2$ denote the corresponding square deviations.
\bigskip

\subsection{Deformed dispersion relation}
\bigskip

Let us now consider a photon moving in this spacetime. Its four--momentum, $p^{\mu} = (E, \vec{p})$, satisfies 

{\setlength\arraycolsep{2pt}\begin{eqnarray}
p^{\mu}p_{\mu} = e^{\psi(x)}\eta_{00}(E)^2 + e^{\zeta(x)}\eta_{ij}(p_ip_j)= 0.  
\end{eqnarray}}

From here on we set $c = 1$, and also define $\phi(x) = \psi(x)- \zeta(x)$. At this point it is noteworthy to mention that the conditions imposed upon the metric fluctuations imply that $<\psi(x)\zeta(x)> = 0$.

From our starting conditions we find that

{\setlength\arraycolsep{2pt}\begin{eqnarray}
p^2 = E^2\Bigl[1 + \phi(x) + O(\phi^2(x))\Bigr].  
\end{eqnarray}}

Keeping only $\phi(x)$ in (9), and averaging, we obtain

{\setlength\arraycolsep{2pt}\begin{eqnarray}
<({p^2 - E^2\over E^2})^2> = \sigma_1^2 + \sigma_2^2.  
\end{eqnarray}}

Let $L$ denote the largest distance between two points, such that they behave in a coherent way under the fluctuations $\zeta(x)$ (while $T$ is the corresponding time associated with $\psi(x)$). If we accept that: (i) these fluctuations are quantum gravity corrections to the Minkowskian metric, and (ii) Planck length might appear together with other length scales in the problem [11]; then we may introduce the following assumption

{\setlength\arraycolsep{2pt}\begin{eqnarray}
\sigma_1^2 = a^2L^2_p/L^2,  
\end{eqnarray}}

{\setlength\arraycolsep{2pt}\begin{eqnarray}
\sigma_2^2 = e^2T^2_p/T^2.  
\end{eqnarray}}

Here $T_P$ and $L_p$ denote the Planck length and time, respectively, while $a$ and $e$ are real numbers. This kind of relation, between square deviation and Planck length, has already been derived [10, 12]. Of course, this Ansatz requires a deeper analysis, and it does not discard other possibilities [13]. Clearly our model contains four free parameter, i.e., $e, a, T$, and $L$, which can not be deduced in the context of our assumptions.

Introducing the real number $\beta = \sqrt{a^2 + b^2}$ (here we have defined $T = \chi L$, and $b = e/\chi$, with $\chi\in Re$), we may rewrite (10) as

{\setlength\arraycolsep{2pt}\begin{eqnarray}
\Bigl({p^2 - E^2\over E^2}\Bigr) = \beta L_p/L.  
\end{eqnarray}}

In other words, we may reduce the number of free paramaters from four to only two, $\beta$ and $L$.

Assuming that the Hamiltonian formulation of classical mechanics ($v = {\partial E\over \partial p}$) is, approximately, valid, we find the speed of propagation of the photon, from here on $c$ is introduced explicitly

{\setlength\arraycolsep{2pt}\begin{eqnarray}
v = c\Bigl[1 - {\beta L_p\over 2L}\Bigr].  
\end{eqnarray}}
\bigskip

\subsection{Generalized Uncertainty Principle}
\bigskip

Let us now consider a simple experiment, the position of the photon will be now monitored. In this situation (and remembering that expressions (13) and (14) have already a statistical meaning, since they have been deduced by an averaging process) we will have an additional uncertainty source, namely, the speed of the photon has an uncertainty given by 

{\setlength\arraycolsep{2pt}\begin{eqnarray}
\Delta v = c{\beta L_p\over 2L}.  
\end{eqnarray}}
\bigskip

Hence, combining linearly the effects of our two uncertainty sources, the uncertainty in the position becomes

{\setlength\arraycolsep{2pt}\begin{eqnarray}
\Delta x = c\Delta t + c{\beta L_p\over 2L}\hat T.  
\end{eqnarray}}
\bigskip

Here $\Delta t$ denotes the uncertainty in the emission of the photon, and $\hat T$ is the time the experiment lasts. Assuming that $\Delta t \sim  \hbar/\Delta E$ (this condition could be not very stringent, indeed, it is valid not only in the context of quantum mechanics, but even in broader frameworks, for instance, in the quantum deformations of the $D = 4$ Poincar\'e groups [14]) we may cast (16) as 

{\setlength\arraycolsep{2pt}\begin{eqnarray}
\Delta x = \hbar/\Delta p + c{\beta L_p\over 2L}\hat T.  
\end{eqnarray}}

The modification of the uncertainty principle is not a new issue [1, 2, 3], and can be derived in the context of quantum geometry [15], black--hole effects [16], quantum measurements at Planck scale [17], and even in Newtonian gravity theory [18]. In the present work we have derived it without considering a particular quantum gravity theory, as loop quantum gravity [7], or resorting to a relation between mass and the radius of a Schwarzschild black--hole [1].

Expression (17) implies that the corresponding Heisenberg algebra is not the usual one.  Its is readily checked that our uncertainty principle can be obtained from the following commutator for the quantum operators $\hat x$ and $\hat p$

{\setlength\arraycolsep{2pt}\begin{eqnarray}
[\hat x, \hat p] = i\hbar\Bigl(1 + c{\beta L_p\over 2\hbar L}\hat T\sqrt{(\hat p - <\hat p>)^2}\Bigr).  
\end{eqnarray}}
\bigskip
\bigskip

\section{Conclusions}
\bigskip

Assuming that quantum gravity corrections appear as nonconformal stochastic fluctuations of the metric it has been shown that a deformed dispersion relation, and a modification of Heisenberg algebra, unavoidably, emerge. 

Concerning our model it is noteworthy to mention that it establishes a peculiar asymmetry. Indeed, it seems to discard the most general case [2], namely, the possibility of having on the right hand side of (18) functions of the position operator, $\hat x$. Expression (8) also allows us to conclude that conformal fluctuations do not lead to a modification of Heisenberg algebra (there is also no deformed dispersion relation).  

From previous results [10] it seems that nonconformal fluctuations would also imply the presence of decoherence effects, i.e., decoherence between macroscopically di\-fferent situations would appear as a consequence of the gravitational vacuum (clearly this statement has to be supported with the corresponding calculations, an issue that will be published elsewhere).

An interesting claim comprises the possibility that the so called $\kappa$--Poincar\'e sy\-mmetries could contain some of the physics of the quantum gravity vacuum [3, 14]. Starting with (17) we may confront our conclusions with the implications of the quantum $\kappa$--Poincar\'e group (here we bear in mind expression (2.13b) of [14]). In order to do this, let us consider a photon with average energy given by the usual expression, $E = \hbar\nu$, then $\nu = c{\beta L_p\over 2\alpha\hbar L}\hat T\Delta p$ (here $\alpha$ is the parameter in (2.13b) of [14]). Imposing energy conservation, we deduce $\Delta p\sim 1/\hat T$. This last fact means that the root--mean--square deviation, $\sigma$, associated with the measurement of distance fulfills $\sigma\sim\hat T$, i.e., a behavior already deduced [9], and which matches with the claim [13] that the dynamics, underlying the fundamental nature of spacetime, does not produce any dissipation, and, in consequence, the fluctuation--dissipation theorem is fulfilled without having a restoring mechanism associated with the intrinsic fluctuations of spacetime. In other words, the matching of our model and the quantum $\kappa$--Poincar\'e group entails the loss of quantum information, though this happens in the present work at a rate faster than in the usual case [9, 13], where this loss grows as $\hat T^{1/2}$. 

From the last remarks we may state that the loss of quantum information is related to the fact that the spacetime symmetries are described by the quantum $\kappa$--Poincar\'e group, and not by the usual Poincar\'e symmetries. An interesting issue in connection with this statement comprises the formulation of a quantum measurement theory employing as spacetime symmetries those of the quantum $\kappa$--Poincar\'e group.

Recently it has been claimed [8] that if a theory predicts that photons propagate with an energy dependent velocity, then Lorentz invariance breaks down. In our case this should be no surprise, since we have shown that we may incorporate in the present model the quantum $\kappa$--Poincar\'e group. Considering a non trivial boost transformation (and also $\beta \not = 0$) along the direction of propagation of the photon, we find that Lorentz invariance breaks down, namely, $\beta /L$ does not behave in the usual way.

We may now evaluate the possible order of magnitude of the extra term in (13). To set a very rough upper bound on these kind of contributions, let us assume that the measurements readouts of the speed of light, $c[1 \pm\epsilon]$, are such that $\epsilon$ (the uncertainty in the experimental results) stems, exclusively, from fluctuations of the metric, namely, we have $\epsilon = {\vert\beta\vert L_p\over 2L}$. This is not a realistic assumption, nonetheless, it will allows us to find a rough upper bound to these effects. From already known results for the speed of light [19] we obtain 

{\setlength\arraycolsep{2pt}\begin{eqnarray}
L/\vert\beta\vert\leq10^{-25}cm.  
\end{eqnarray}}
\bigskip
\bigskip

\Large{\bf Acknowledgments.}\normalsize
\bigskip

The author would like to thank A. A. Cuevas--Sosa for his 
help. 

\end{document}